\documentclass{aastex}
\usepackage{emulateapj5}

\newif\ifAMStwofonts
\AMStwofontstrue

\def\kms{km~s$^{-1}$}

\def\degree{$^{\circ}$}

\def\ga{\mathrel{\hbox{\rlap{\hbox{\lower4pt\hbox{$\sim$}}}\hbox{$>$}}}}
\def\la{\mathrel{\hbox{\rlap{\hbox{\lower4pt\hbox{$\sim$}}}\hbox{$<$}}}}

\shorttitle{HI kinematics in the SMC}

\shortauthors{S.\ Stanimirovi\'{c} et al.}

\begin{document}

\title{A New Look at the Kinematics of Neutral Hydrogen in 
the Small Magellanic Cloud}

\author{S.\ Stanimirovi\'{c}}
\affil{Radio Astronomy Lab, UC Berkeley, 601 Campbell Hall, Berkeley, CA 94720}
\email{sstanimi@astro.berkeley.edu} 
\author{L.\ Staveley-Smith}
\affil{Australia Telescope National Facility, CSIRO, P.O.\ Box 76, 
Epping, NSW 1710, Australia}
\author{P. A.\ Jones}
\affil{Australia Telescope National Facility, CSIRO, P.O.\ Box 76, 
Epping, NSW 1710, Australia}
\submitted{To appear in ApJ, 2004 March 20}

\begin{abstract}
We have used the latest HI observations of the Small
Magellanic Cloud (SMC), obtained with the Australia Telescope 
Compact Array and the
Parkes telescope, to re-examine the kinematics of this dwarf, irregular galaxy.
A large velocity gradient is found in the HI velocity field with a 
significant symmetry in iso-velocity contours, suggestive of a 
differential rotation. A comparison of HI data with the predictions from
tidal models for the SMC evolution suggests that the central region of the
SMC corresponds to the central, disk- or bar-like, component left from 
the rotationally supported SMC disk prior to its last 
two encounters with the Large Magellanic Cloud. In this scenario, the
velocity gradient is expected as a left-over
from the original, pre-encounter, angular momentum. 
We have derived the HI rotation curve and the mass model for the SMC. 
This rotation curve rapidly rises to about 60 \kms~up to the turnover 
radius of $\sim3$ kpc. A stellar mass-to-light ratio of about
unity is required to match the observed rotation curve, suggesting that a
dark matter halo is not needed to explain the dynamics of the SMC. 
A set of derived kinematic parameters agrees
well with the assumptions used in tidal theoretical models that led to a good
reproduction of observational properties of the Magellanic System. The
dynamical mass of the SMC, derived from the rotation curve, 
is $2.4\times10^{9}$
M$_{\odot}$. 
\end{abstract}


\section{Introduction}

The Small Magellanic Cloud (SMC) is a nearby\footnote{Throughout this paper
we assume a distance to the SMC of 60 kpc \citep{Westerlund97}.}, 
gas-rich,  dwarf irregular galaxy.
Its morphology, dynamics, and evolution are highly complex, and must 
have been heavily influenced by gravitational 
interactions with the nearby Large Magellanic Cloud (LMC) and the Galaxy 
\citep{Mary-nature}. As a dwarf irregular galaxy, the SMC is different 
from our own Galaxy in many respects, having a lower heavy element abundance,
significantly lower dust content, and a consequently 
stronger interstellar radiation field \citep{Stanimirovic99,Stanimirovic01}.

Many recent high resolution studies have shown that dwarf galaxies in
general have a very dynamic interstellar medium (ISM), 
structured mainly by  star formation and its aftermath. The most 
obvious  example of star-formation  creativity are numerous expanding 
shells of gas \citep{Puche92,Staveleyetal97,Kim99,Stanimirovic99,Walter99}. 
Dwarf galaxies are also usually found to have large, and dynamically 
important, halos of dark matter \citep{Mateo98}. 
The relative roles of a galactic rotation, pressure support, dark matter, 
and magnetic fields on the 3-D structure and dynamics of these galaxies are
still open questions. 
Furthermore, it is not clear yet whether
different stages of galactic evolution are marked by fundamentally 
different phenomena, and whether these phenomena are selective with 
respect to spatial scales.
 
It has been known for some time that the morphology and kinematics of 
the SMC traced by different stellar populations show very different
properties \citep{Gardiner92}. This has been one of the major reasons 
to start thinking about forces other than gravitational that may have 
played a significant role in the formation and evolution of the whole 
Magellanic System.  The questions of morphology and kinematics are 
closely related to the long-standing and greatly controversial 
questions of the SMC's 3-D structure and depth along the line-of-sight.
As the results of several new optical and near-IR surveys are becoming
available \citep{Zaritsky00,Cioni00,Maragoudaki01},
revealing structural evolution of the SMC, as well as new N-body simulations of
the dynamical evolution of the SMC \citep{Yoshizawa03}, 
we find that it is timely to re-examine the morphology and kinematics 
of the SMC as traced by the neutral hydrogen (HI).

The mapping of HI in the SMC has a long and productive history. After the
pioneering work by \citet{KerrHindmanRobinson54} and
\citet{Hindmanetal63}, \citet{Hindman67} was the first to notice the
velocity gradient in the SMC and to model its rotation curve, followed by
\citet{BajajaLoiseau82}. However, the velocity field of the SMC is far
from a simple text-book example.  The HI profiles, often complex and with
multiple peaks,  have caused much controversy in the past, being
interpreted as due to either expanding shells of gas or spatially separate systems
\citep{Hindman67,MathewsonFordVisvanathan88,Martin89}.  An improvement
by a factor of ten in spatial resolution, relative to previous surveys with
the Parkes telescope, was achieved in the HI survey by
\citet{Staveleyetal97}, using the Australia Telescope Compact Array
(ATCA). These new high resolution data showed that much of the HI profile
complexity lies in the huge number ($\sim$ 500) of expanding shells.
The new high resolution HI observations were complemented with new low resolution
observations obtained with the Parkes telescope to provide information over
a continuous range of spatial scales from 30 pc to 4 kpc \citep{Stanimirovic99}.

The aim of this paper is to re-examine the HI kinematics of the SMC, 
as viewed from high resolution observations, and compare it with predictions from 
tidal theoretical models.  We start by summarizing the HI observations in 
Section~\ref{s:hi-data}. 
The morphology and kinematics of the SMC from the HI distribution, as well as
viewed using other tracers, are discussed
in Section~\ref{s:overview}.
The SMC 3-D structure and line-of-sight depth are reviewed briefly in
Section~\ref{s:3D-structure}. We then compare HI data with predictions
from several tidal models in Section~\ref{s:theoretical-models}.
The rotation curve and mass model of the SMC are investigated in 
Section~\ref{s:rotation-curve}. Finally, a summary and concluding remarks
are given in Section~\ref{s:summary}.

\section{HI data}
\label{s:hi-data}

\begin{table*}
\caption{\label{t:summary} Radio and optical properties of the SMC.}
\centering
\begin{tabular}{lc}
\noalign{\smallskip} \hline \hline \noalign{\smallskip}
Property& Value\\
\hline 
    \\
{\bf Radio:}& \\
{\it Measured:}\\
RA (J2000)$^{a}$  & 01$^{\rm h}$ 05$^{\rm m}$ \\
Dec (J2000)$^{a}$ & $-72^{\circ}$ $25'$ \\
Systemic velocity$^{b}$, $V_{\rm sys}$ (Gal.) & 24 \kms \\
Systemic velocity$^{b}$, $V_{\rm sys}$ (Hel.) & 160 \kms \\
HI mass$^{c}$, $M_{\rm HI}$ & $4.2 \times 10^{8}$ M$_{\odot} $\\

{\it Model-dependent:}\\
Inclination$^{d}$, $i$ & 40\degree $\pm$ 20\degree\\
Kinematic Position angle$^{d}$, PA & 40\degree \\
V$_{\rm max}^{d}$ & 60 \kms \\
R$_{\rm max}^{d}$ & 2.5 -- 3 kpc \\ 
M$_{\rm dyn}^{d}$     & $2.4\times10^{9}$ M$_{\odot}$ \\
    \\

{\bf Optical: }&    \\
Extinction$^{e}$, $E_{\rm B-V}$  & $\sim0.05$--0.25 mag\\
Total visible magnitude$^{f}$, $V_{\rm T}$  & $+2.4$ mag\\
Absolute visible magnitude$^{f}$, $M_{\rm V}$ & $-16.5$ \\
Visible luminosity$^{f}$, $L_{\rm V}$ & $3.1\times10^{8}$ L$_{\odot}$ \\
$M_{\rm H}/L_{\rm V}$ & 1.4 M$_{\odot}$/L$_{\odot}$ \\
$(B-V)_{\rm T}^{0}$ & 0.41 \\
$(U-B)_{\rm T}^{0}$ & $-0.23$ \\

\noalign{\smallskip} \hline \noalign{\smallskip}
\end{tabular}
\tablenotetext{a}{The apparent kinematic center derived as the 
intersection of the principal kinematic axes.}
\tablenotetext{b}{Systemic velocity of the apparent kinematic center
derived from the velocity field.}
\tablenotetext{c}{\citet{Stanimirovic99}.}
\tablenotetext{d}{Derived from the tilted ring analysis 
in Section~\ref{s:rotation-curve}.}
\tablenotetext{e}{\citet{Zaritsky02}.}
\tablenotetext{f}{\citet{RC3}.}
\end{table*}

\begin{figure*}
\caption{\label{f:HI-density} {\bf [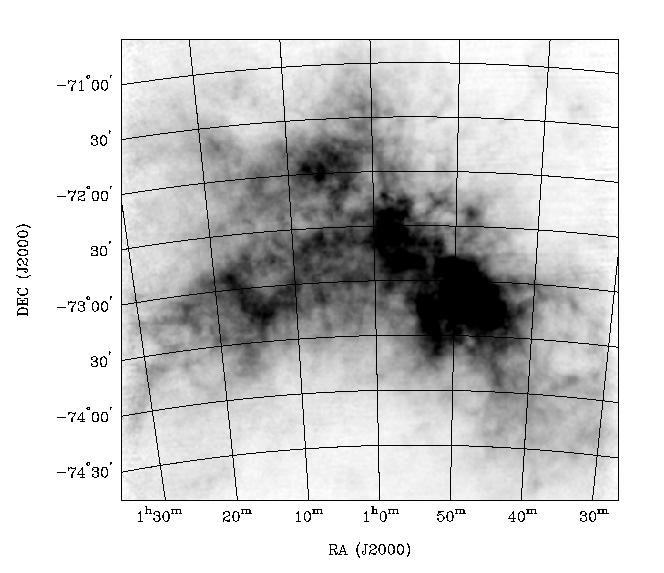]} An HI 
column-density image of the SMC. The
grey-scale intensity range is 0 to $7 \times 10^{21}$ atoms cm$^{-2}$
with a linear transfer function. The maximum HI column density, $1.43
\times 10^{22}$ atoms cm$^{-2}$, is at position RA $00^{\rm h} 47^{\rm m}
33^{\rm s}$, Dec $-73^{\circ} 05' 26''$ (J2000).}
\end{figure*}

The HI in the SMC was observed with the ATCA\footnote{The Australia 
Telescope is funded by the Commonwealth of Australia for
operation as a National Facility managed by CSIRO.}, a radio interferometer, 
in a mosaicing mode (Staveley-Smith et al. 1997). 
Observations of the same area were obtained also with the
64-m Parkes telescope. The two sets of observations were then combined
(Stanimirovi\'{c} et al. 1999), resulting in the final HI data cube with
angular resolution of 98 arcsec, velocity resolution of 1.65 \kms, and
1--$\sigma$ brightness-temperature sensitivity of 1.3 K to the full range 
of spatial scales between 30 pc and 4 kpc.
The area covered with these observations is RA $00^{\rm h} 30^{\rm m}$ 
to $01^{\rm h} 30^{\rm m}$
and Dec $-71^{\circ}$ to $-75^{\circ}$ (J2000), over a velocity range of 90
to 215 \kms. For details about the ATCA and Parkes observations, data
processing, and data combination (short-spacings correction) see 
Staveley-Smith et al. (1997) and Stanimirovi\'{c} et al. (1999).

\section{Structure and kinematics of the SMC}
\label{s:overview}

\subsection{HI distribution}

Fig.~\ref{f:HI-density} shows the integrated HI column-density
distribution. The large-scale HI morphology of the SMC is quite irregular and 
does not show obvious symmetry.
The most prominent features are the elongation from the north-east to the
south-west and the V-shaped concentration at the east 
(RA 01$^{\rm h}$ 14$^{\rm m}$, Dec $-73^{\circ}$ 15$'$, 
J2000). These are usually referred to as the `bar' and the Eastern Wing, 
although their dynamical importance is still not well understood. 
A `bridge' of gas 
appears to connect the `bar' and the Wing (note that this is not the
Magellanic Bridge that connects the SMC and the LMC), while 
the arm-like extension of the `bar' towards the
north-east is also prominent. On smaller scales and looking at different
velocity channels (see Stanimirovi\'{c} et
al. 1999), the HI distribution appears very complex and frothy, being
dominated by numerous expanding shells, filaments and arcs. 
The total estimated mass of the HI, after correction for
self-absorption, is $4.2 \times 10^{8}$ M$_{\odot} $ (Stanimirovi\'{c} et
al. 1999). A summary of radio and optical properties of the SMC is given in
Table~\ref{t:summary}.

\begin{figure*}
\caption{\label{f:v-HI} {\bf [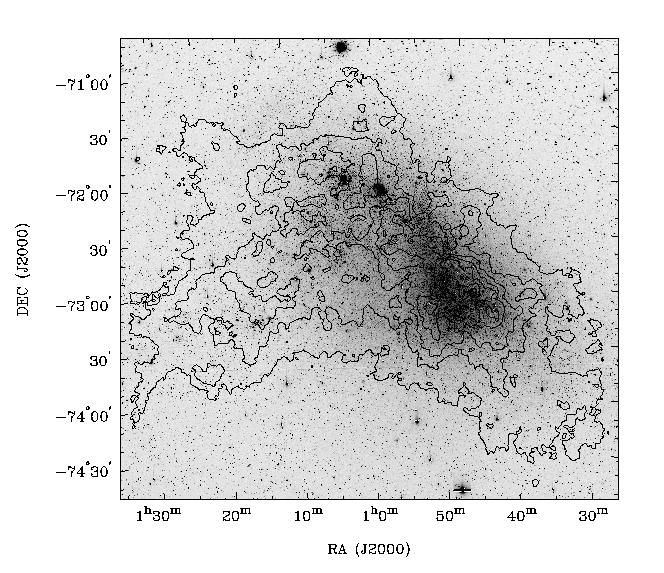]} The HI 
column density image super-imposed on the
V-band CCD image. The contour levels range from $2.0\times10^{20}$ to 
$1.3\times10^{22}$ atoms cm$^{-2}$, with a contour interval of 
$3.8\times10^{20}$ atoms cm$^{-2}$. The V-band image was obtained by Mike
Bessell using the 16$''$ telescope at the Mount Stromlo Observatory.}
\end{figure*}

The HI column density distribution is super-imposed on the V-band  optical
image of the SMC, kindly provided to us by M. Bessell,  in
Fig.~\ref{f:v-HI}. In general, in the `bar' and the Eastern  Wing 
the stellar and HI distributions correlate
well, however the HI is significantly more extended towards the
south-west, the south-east, and the north-west.

\subsection{HI kinematics}
\label{s:profile-analysis}

\begin{figure*}
\caption{\label{f:1mom-mean} {\bf [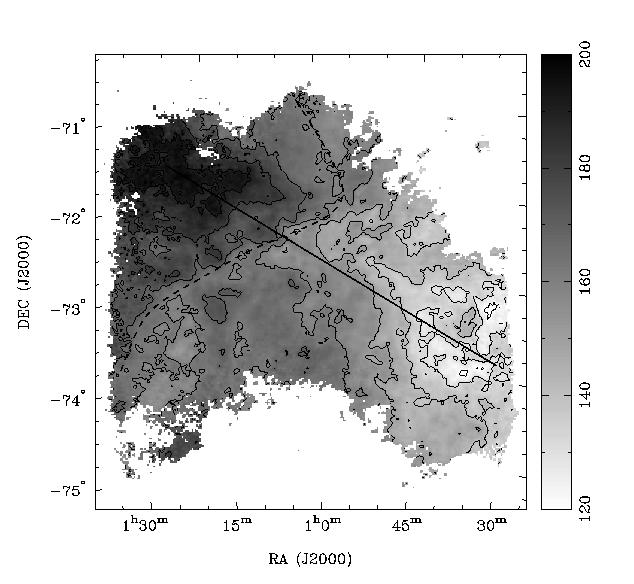]} The 
intensity-weighted mean heliocentric 
velocity field of the
SMC, derived from the combination of the ATCA and Parkes data.  The grey
scale range is 120 to 200 \kms~with a linear transfer function. The contour
levels range from 120 to 200 \kms, with a contour interval of 10
\kms. All pixels with a corresponding HI column density lower than
$9\times10^{20}$ atoms cm$^{-2}$ were masked out. The direction of the largest
velocity gradient is shown with a solid line, and the direction of the
largest velocity distortions, forming a S-shaped features, is shown with
a dashed line. }
\end{figure*}

The intensity-weighted mean velocity along each line-of-sight, or the first
moment map, was derived from the  full resolution HI data cube (corrected
for short spacings), and is shown in Fig.~\ref{f:1mom-mean}.  A large 
velocity gradient, from 91 \kms~in the south-west to 200 \kms~in the
north-east, can be seen.  The iso-velocity  contours show some  large-scale
symmetry suggestive of differential rotation for the main gaseous body of
the SMC. However, clear distortions are visible in the north-west,
most likely corresponding to several shells and filamentary features
aligned into a chimney-like structure at velocity $\sim$ 123 \kms~(Fig. 2
in Stanimirovi\'{c} et al. 1999), and in the south-east,  towards the Eastern
Wing region, where again a supergiant shell  was found 
(494A; see Stanimirovi\'{c} et al. 1999).  
These perturbations form a S-shaped feature
perpendicular to the direction of  the main velocity gradient (as shown in
Fig.~\ref{f:1mom-mean} with a dashed line).

\begin{figure*}
\epsscale{1.}
\plotone{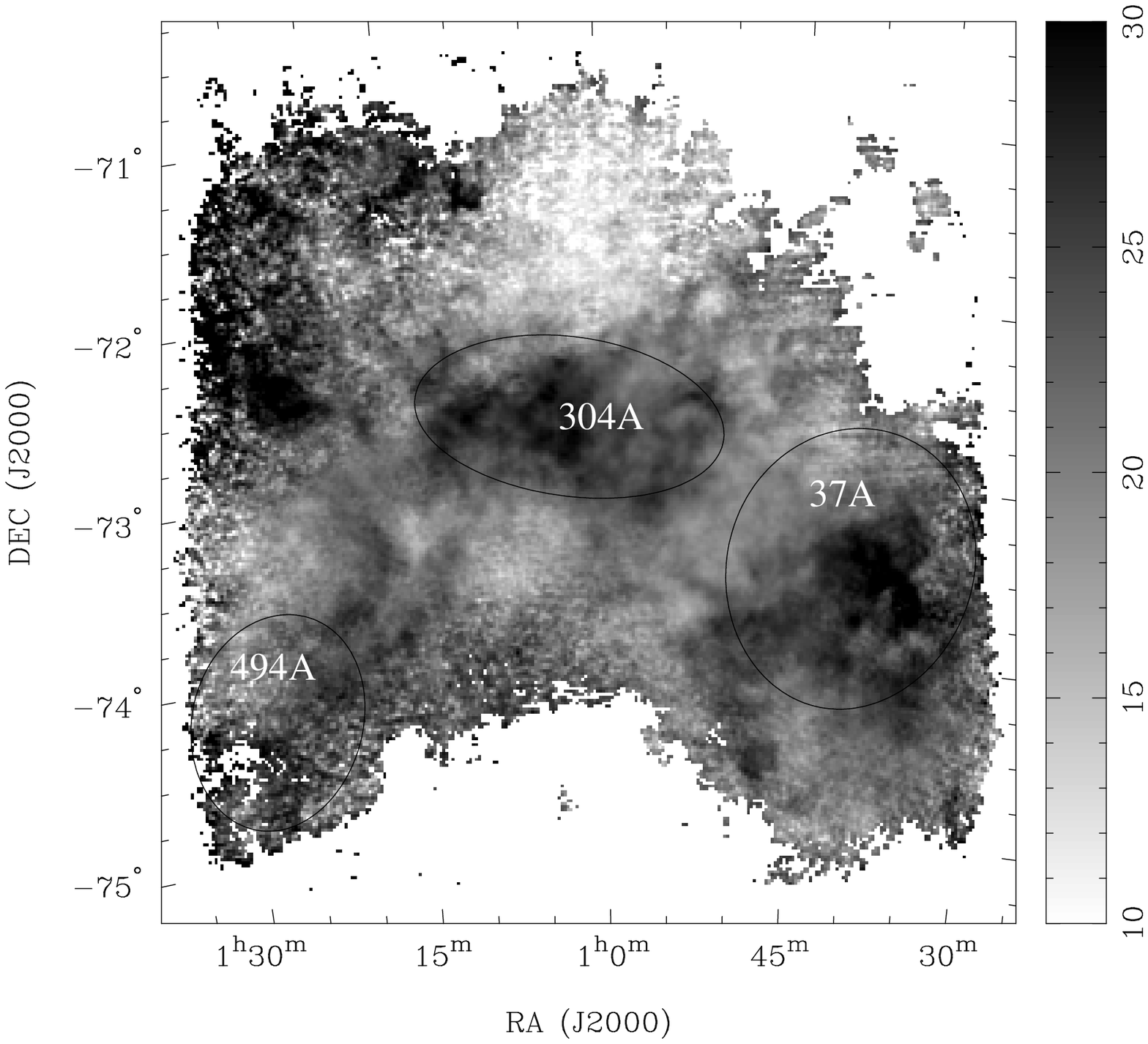}
\caption[The velocity dispersion.]  {\label{f:dispersion-mean} The velocity
dispersion of the SMC from the second moment analysis obtained from the
combined ATCA and Parkes data cube. The grey-scale range
is 6 to 37 \kms. Positions of the three largest shells (37A, 304A and
494A)  from Stanimirovi\'{c} et al. (1999) are overlaid. All pixels with 
a corresponding HI column density lower than
$7\times10^{20}$ atoms cm$^{-2}$ were masked out.}
\end{figure*}

The second moment map or the intensity-weighted velocity dispersion 
is shown in Fig.~\ref{f:dispersion-mean}.
This velocity dispersion varies from $\sim$5 to 40 \kms~across the SMC.
The regions with higher dispersion appear to be associated with 
the positions of the three largest supergiant shells (SGSs). The region
around RA $01^{\rm h}$ $00^{\rm m}$, Dec $-71^{\circ} 30'$ (J2000) has the
lowest dispersion with a mean value of $\sim$10 \kms. As this region 
corresponds to the
receding hemisphere of the SGS 304A  (Stanimirovi\'{c} et al. 1999), the lower
velocity dispersion may be explained with the fact that  most of the
approaching hemisphere of SGS 304A is missing at the northside.

We have investigated two effects that could significantly influence
the observed velocity field shown in Fig.~\ref{f:1mom-mean}: 
(i) the effect of the proper motion of the SMC, and 
(ii) the effect of multiple-peak velocity profiles. \\
(i) The binary motion of the SMC
around the Galaxy, combined with its large angular extent,  could have a
significant contribution to the observed velocity field, as was shown for the
case of the LMC by \citet{LuksRohlfs92} and \citet{Kim98}. This contribution
consists of the projection of the transverse velocity of the SMC's center
of gravity on the line-of-sight at each position. In the case of a large
angular extent on the sky, this projection can vary greatly across the
field of observation and cause a significant change in the observed 
velocity field.
To correct for this effect, we used values for the SMC's proper motion
and the heliocentric transversal velocity predicted by numerical
simulations by \citet{Gardiner94}, which are in agreement with the
estimates based on a combination of proper motion measurements by 
\citet{Kroupa94} and \citet{KroupaBastian97}.
At any position across the SMC the proper-motion corrected heliocentric velocity 
was obtained by subtracting the projection of the heliocentric
transverse velocity of the center of the SMC on the line-of-sight from the
observed radial heliocentric velocity.
The values for the conversion between the observed and the proper-motion 
corrected heliocentric velocity range between $-23$ \kms, 
in the north-west, and 28 \kms, in the south-east. 
The correction for the proper motion, however, did not make as big
difference as in the case of the LMC \citep{LuksRohlfs92}. 
The reason is that the angular extent of the SMC is less than half that 
of the LMC. In addition, the direction of the SMC's motion is 
perpendicular to the observed velocity gradient 
resulting in the gradient being almost unaffected by the proper motion.

(ii) Velocity profiles in the SMC are usually very 
complex, having double or even multiple-peak components.  
Since the intensity-weighted mean velocity (Fig.~\ref{f:1mom-mean})
is biased towards the velocity component with higher intensity, it is not
necessarily the best representation of complex velocity profiles.  To avoid
this bias, we found, for each line-of-sight, the minimum and maximum
velocity for which the intensity is 5\% of the peak value for that 
line-of-sight.  The mean of the minimum and maximum velocity, determined in this
way, was then taken to be our new velocity estimate.  Since such an analysis
requires a high signal-to-noise ratio, we only used the low resolution
Parkes data cube. The resultant velocity field has 
a more regular `spider' pattern than the mean one,  however
irregularities are still present in the north and close to the Eastern
Wing region. 
We also investigated the existence of a bimodal velocity field, 
by modeling each velocity profile in the Parkes data cube with one or a 
superposition of two independent Gaussian functions. As a result, two 
separate velocity structures were depicted, with central velocities corresponding
to the low and high velocity entities of 137 \kms~and 174 \kms,
respectively. The position-velocity diagrams show almost parallel velocity
fields of these two components. However, in many cases the velocity fields
intersect suggesting that two separate velocity components may be a
consequence of the statistical data handling of complex HI profiles with no
substantial physical meaning.

\begin{figure*}
\epsscale{1.}
\plotone{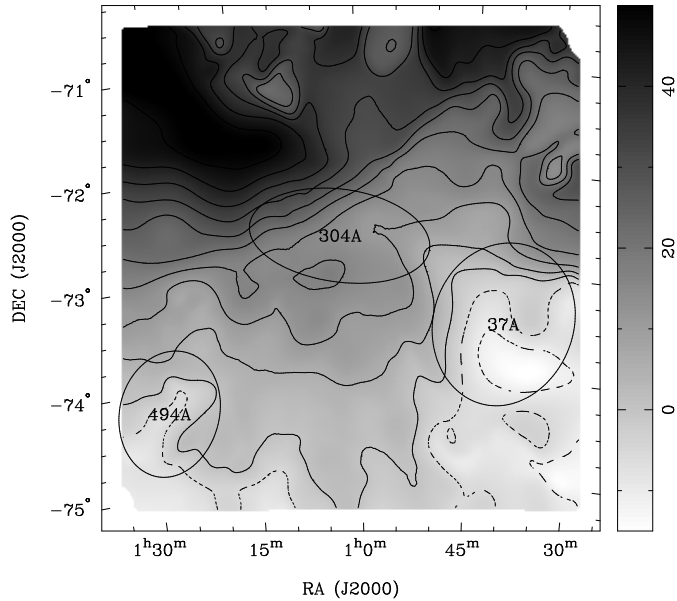}
\caption[The unimodal galactocentric  velocity field corrected for the
proper motion.]{\label{f:gal_velfield} The galactocentric velocity
field of the SMC, from the Parkes HI observations, with angular resolution of
$\sim15$ arcmin, corrected for the proper motion. The grey-scale range is
$-15$ to 50 \kms with a linear transfer function. The contour levels range
from $-20$ to 50 \kms, with an interval of 5 \kms.  The three largest
supershells (37A, 304A and 494A) from Stanimirovi\'{c} et al. (1999) are
superimposed.}
\end{figure*} 

The proper-motion corrected velocity field in the 
Galactocentric velocity reference frame, derived from the Parkes HI data
cube using the 5\% cut-off described in (ii), is shown in 
Fig.~\ref{f:gal_velfield}.
A clear velocity gradient is noticeable from $-15$ \kms,
in the south-west, to 53 \kms, in the north-east. The positions of the three
largest shells in the SMC are also overlaid in Fig.~\ref{f:gal_velfield},
suggesting that supershells 304A and 494A may still be responsible  for
some of the perturbations in the velocity field. Some perturbations are
also visible in the north-west.

\subsection{Comparison with other tracers}
\label{s:other-tracers}

\subsubsection{SMC morphology from stellar populations}

It has been known for some time that 
young and old stellar populations in the SMC have significantly different
spatial distributions (e.g. \citet{Gardiner92}).
Several recent stellar surveys have especially enhanced our knowledge about
the different morphological properties of different stellar populations.
The recent stellar survey of the central 
SMC (4\degree $\times$ 4\degree)  by \citet{Zaritsky00} showed that 
while young stars
(with ages $\la 200$ Myr) exhibit an irregular distribution, similar
to that seen in HI, the old stellar population (with ages $\ga 1$ Gyr)
shows regular, undisturbed, and almost spheroidal distribution.  
Similar results were reached from the {\sc DENIS} survey by \citet{Cioni00}
based on $IJK_{s}$ IR bands. These studies also pointed out that
the `bar' and the Eastern Wing region may not be distinct dynamical entities,
but could be a result of recent hydrodynamic interactions between 
the LMC and SMC's gaseous components. 
\citet{Maragoudaki01} further investigated the dynamical origin of the `bar'
by using isodensity contour maps of stars with different ages. They found 
similar results concerning old stellar populations. In addition, 
their data show that the `bar' and the Eastern Wing region were already 
prominent features some 0.3--0.4 Gyr ago; however, it is not clear whether 
this is a temporary extended region of star formation or a genuine,
dynamical feature.

\subsubsection{SMC kinematics from stellar populations}
\label{s:other-tracers-kinematics}

The intermediate age and old stellar populations, traced by 
planetary nebulae \citep{Dopita85b} and carbon stars
\citep{Hardy89,Hatzidimitriou97,Hatzidimitriou99,Kunkel00}, do not show  signs of
rotation. The red horizontal branch clump stars in the north-east 
show a velocity gradient of 70 \kms~which was interpreted as due to 
a distance spread of 10 kpc along the line-of-sight (Hazidimitriou et al. 1993).  
The young stellar population, traced by HI shells, shows a clear velocity
gradient from the south-west to the north-east \citep{Staveleyetal97},
while Cepheids may have a similar trend with velocity
\citep{MathewsonFordVisvanathan88}. 
All populations show similar mean heliocentric velocity and velocity
dispersion regardless of age and location
\citep{Hatzidimitriou99,Staveleyetal97}.
 
From an analysis of 150 carbon stars, \citet{Kunkel00} estimated the inclination
of the SMC orbital plane to $i=73^{\circ}\pm4^{\circ}$ relative to the
plane of the sky. They found that carbon stars are associated with 
more negative velocity and denser HI, while low density HI seems to be 
devoid of carbon stars. This suggests that non-gravitational 
forces must be acting on 
the gaseous component to provide its separation from stellar systems.
Similar estimates for the SMC's inclination were reached using Cepheids by
\citet{Caldwell86}, $i=70^{\circ} \pm 3^{\circ}$, and by
\citet{Groenewegen00}, $i=68^{\circ} \pm 2^{\circ}$.

\section{The 3-D structure of the SMC}
\label{s:3D-structure}

\subsection{Previous observational models}

The 3-D structure of the SMC has been a matter of great controversy in the
past. Since early mapping of HI, complex HI profiles pointed to the
existence of unusual motions of the gas \citep{Hindman67}. 
\cite{Hindman67} also suggested the first model of
the SMC, as a flattened disk with three supergiant shells in the main body.

From the analysis of the radial velocity distributions of HI, stars, HII
regions, and planetary nebulae, \cite{MathewsonFord84} and Mathewson, Ford
\& Visvanathan (1986, 1988) suggested the `two separate entities' model for
the SMC, whereby the SMC is, along most of its angular extent, broken into
two velocity subsystems (the Small Magellanic Cloud Remnant, SMCR, and the
Mini Magellanic Cloud, MMC). In this model, 
the SMCR and MMC are separated in velocity by
$\sim$ 40 \kms, and have their own nebular and stellar populations.  The
reason proposed for this disruption of the SMC was its close
encounter with the LMC some 2$\times$10$^{8}$ yr ago. 
A slight modification of the `two separate entities' 
model was suggested by \cite{Torres87}, and \cite{Martin89}, who found four, 
instead of two, different velocity components.

\cite{Caldwell86} also used Cepheids to suggest the `bar and three arms'
geometrical model of the SMC. 
The central bar is very elongated (5-to-1) and is seen edge-on.
A distant arm of material is pulled from the center of the SMC to 
the west, while two sides of the bar located in the north-east and 
the south-west are considered as two near arms.

\subsection{Depth along the line-of-sight}
\label{s:smc-depth}

The question of the 3-D structure of the SMC is closely related to the
long-standing and greatly controversial issue of the SMC's depth along the
line-of-sight. By measuring distances and radial velocities of 161 Cepheids, 
\cite{MathewsonFordVisvanathan86} found a great depth along the
line-of-sight of $\sim$30 kpc, and a distance gradient from the north-east to
the south-west. The large depth of the SMC was supported later by
\cite{MathewsonFordVisvanathan88}, using Cepheids again, and by 
\cite{Hatzidimitriou89} and \cite{Hatzidimitriou93} from a study 
of the intermediate-age (halo) population. 

\cite{Welch87} have pointed out that the determination of the Cepheid 
 distances in
\cite{MathewsonFordVisvanathan86} suffers from a few possible problems: 
an inconsistent correction to mean magnitude, the assumption of zero
intrinsic scatter for the period-luminosity (P-L) relation, and 
sample inhomogeneity.  \cite{Welch87} concluded
that the SMC does not extend in depth beyond its tidal radius (4 -- 9 kpc).
\cite{Martin89} confirmed that the young stars in the SMC lie within a depth
of $<10$ kpc.
On the other hand, Groenewegen (2000) used near-infrared observations
of Cepheids, which are less affected by reddening
and metallicity, and derived the depth of the SMC of 14 kpc, assuming
that the P-L relation has the same scatter for both the LMC and the SMC.

Although Cepheids are used as the main distance indicators, the P-L relation
requires an independent zero-point calibration, which is very difficult 
to achieve, and could suffer from a significant metallicity dependence. 
Another source of error in the distance determination, that has not been
previously fully appreciated, is differential extinction toward the SMC.
While reddening of Cepheids in the SMC has been extensively studied (see
\cite{Welch87} for a summary), a constant value has been often
applied for distance determination.
For example, \cite{MathewsonFordVisvanathan86} found no sign of
differential reddening and assumed a uniform correction of 0.06 mag for
their sample of 161 Cepheids. 

Recently, \cite{Zaritsky02} determined the extinction map across the SMC,
using the Magellanic Clouds Photometric Survey, and showed that 
extinction varies both spatially across the SMC and with stellar
population. They found that young, hot stars have an average extinction 
0.3 mag higher than old, cool stars, and that young stars 
show a significant increase in extinction along the main SMC ridge,
from the north-east to the south-west. 
\cite{Zaritsky02} found $E_{\rm B-V} \sim$ 0.05 -- 0.25 mag.
For comparison, the mean value for the interstellar reddening previously 
measured by \cite{Caldwell85} was $E_{\rm B-V}=(0.054 \pm 0.021)$
mag, while \cite{Sasselov97} found a slightly higher value of 
$E_{\rm B-V} = (0.125 \pm 0.009)$ mag.

The differential extinction across the SMC has a significant influence 
on the distance estimate. 
Differential reddening toward Cepheids in the south-west part of the SMC is 
$\sim0.2$ mag higher than toward in the north-east.
This results in the interstellar infrared absorption being 
larger in the south-west by $\sim$0.4 mag. 
Hence, the Cepheid distances estimated by \cite{MathewsonFordVisvanathan88} 
could be over-estimated by up to 18\%. This corresponds to 
about 10 kpc at a  distance of 60 kpc, and suggests that 
the correction for the interstellar absorption can significantly 
influence the distance determination, and
easily bring the depth of the SMC within its tidal radius (4 -- 9 kpc).

\section{Comparison with theoretical models}
\label{s:theoretical-models}

There are two families of theoretical models, based on either a tidal 
or a ram pressure scenario, that try to reproduce observational 
features in the Magellanic System, caused by interactions between 
the SMC, LMC, and the Galaxy. 
The 3-D structure and kinematics of the SMC were particularly addressed 
in tidal models by \citet{Gardiner94}, \citet{Gardiner96}, and 
\citet{Yoshizawa03}. The recent work by \citet{Yoshizawa03}
provides, to date, the most detailed model for the evolution of the SMC.
We summarize here results from these tidal simulations and 
their major predictions concerning the structure and kinematics of the SMC.

It is important to note that none of the theoretical models so far predicts 
the bimodal velocity distribution throughout the main gaseous body of the SMC.
This gives support to the idea that most, if not all,
of the observed line-splitting comes from the combined effects of 
numerous expanding shells (Staveley-Smith et al. 1997). 

\subsection{Model predictions}

The above three models come from N-body simulations of the gravitational 
interactions in the Galaxy-LMC-SMC system to reproduce the 
observed gas distribution in the Magellanic System, primarily those of the
Magellanic Stream and Bridge. While \citet{Gardiner94} 
modeled the SMC as a single-component disk-like system, 
in simulations by \citet{Gardiner96} the SMC was represented by a 
two-component particle system consisting of a nearly spherical halo and a
rotationally supported disk, with a disk-to-halo mass ratio of 1:1. 
This high mass ratio resulted in the original disk quickly becoming
unstable, and being transformed into a bar-like structure.
The tidal model by \citet{Yoshizawa03} included, for the first time, 
gas dynamics and  star formation processes, while representing the SMC 
also as a rotationally supported exponential disk with a nearly spherical 
halo, having the disk-to-halo mass ratio of 3:7.
The significantly lower disk mass resulted in the disk being
stable against bar instabilities.
Following typical findings for the Magellanic-type galaxies,
a slowly rising rotation curve was assumed in both models, with a turnover 
radius of 3.5 kpc and maximum rotation velocity of 50 \kms.
The best spatial orientation of the original SMC disk was determined in
\citet{Gardiner96} to have an inclination $i=45^{\circ}$ and
major axis $\theta=230^{\circ}$ in order to match the observed 
gas and stellar distributions.
These values were adopted by \citet{Yoshizawa03} and
simulations were repeated for different star formation parameters.

The major and common result of all three models is that the 
current 3-D structure of the SMC is composed of a central, disk- or bar-like, 
component and two tidal, spiral-arm-like
tails. These tidal tails extend into the Magellanic Bridge, were formed
200 Myr ago, and are seen in both gas and stellar components. 
In particular, simulations by \citet{Yoshizawa03}
show great morphological change of the initial SMC's gas disk --
from its original size of about 10 kpc in diameter the disk
shrunk forming first the Magellanic Stream and the Leading Arm 
(about 1.5 Gyr ago), and then later two tidal arms 
that form the Magellanic Bridge (about 200 Myr ago). 
At the end of the simulation the gas disk is still present but
is now significantly smaller in size, approximately by a factor of 2.5. 
The current SMC consists of the following components.
\begin{enumerate}
\item The left-over {\bf disk-like component}, located at 
0$^{\rm h} 15^{\rm m}<$RA$<1^{\rm h}$, is not
greatly elongated along the line-of-sight, with the stellar disk having a higher
distance dispersion than the gas disk. 
Kinematically, the gas disk has a significant
velocity gradient, from $\sim$80 to 180 \kms. 
The left-over stellar disk, on the other hand, shows high dispersion in the
velocity field but no significant velocity gradient. 
In simulations by \citet{Gardiner96}, this is a bar-like feature 
slightly bigger and 
centered more westward, covering 0$^{\rm h} 30^{\rm m}<$RA$<1^{\rm h} 30^{\rm m}$,
and a slightly wider velocity range.

\item The {\bf eastern tail}, starting around RA $\sim1^{\rm h}$, extending into
the Magellanic Bridge, and covering a distance range from 55 to 40 kpc. This
feature starts at a heliocentric velocity of about 170 \kms~and gradually
decreases to about 120 \kms~at RA $>2^{\rm h}$.
It is seen in both gas and stars but the gas tail is more extended.

\item The {\bf western tail}, starting around RA $\sim0.5^{\rm h}$ and covering
a distance range from 55 to almost 80 kpc. Kinematically, this feature starts
westward from the main disk-like component at a velocity of 
about 80 \kms, but then turns east
passing the main component with increasing velocity all the way to about
$>250$ \kms~around RA $>2^{\rm h}$. The stellar western tail follows the gas one
and is less extended.  
\end{enumerate}
Hence, the two tidal tails contribute mostly to the SMC's 
elongation along the line-of-sight, while the SMC disk-like component is 
not significantly elongated ( $\sim$5 kpc).

The best simulation by \citet{Yoshizawa03} reproduces a large number of
observational properties in the Magellanic System, suggesting that 
most of them may be predominantly of tidal origin. 
Very importantly, a good morphological and kinematic reproduction 
of the Magellanic Stream with almost no stars was achieved for the first time,
by using a very compact initial configuration of the SMC's stellar disk.  
The major shortcoming of the model is
its failure to reproduce gas masses and mass ratios in the 
Magellanic Stream, Bridge, and the SMC. This may be related with the 
need for a more massive initial disk of the SMC, 
or a different halo-to-disk mass ratio, and requires further investigation.
In addition, as pointed out by \citet{Putman03}, a number of detailed
features related to the Magellanic Stream (e.g. its double-helix
appearance and numerous small clouds surrounding its main filaments) 
need to be explained. It has also been shown that
forces other than gravity must play a significant role in the kinematics 
of the SMC \citep{Kunkel00,Zaritsky00}.

The results of tidal numerical simulations show that the 
current central gaseous body of the SMC ($\sim4$ degrees in extent)
has a significant velocity gradient. \citet{Gardiner96} explained this
gradient as being due to the elongation of the SMC's bar-like component 
along the line-of-sight.
However, a significant angular momentum left from before the last 
two close encounters with the LMC is expected to be present, if the
pre-encounter SMC disk had an internal spin. 
The N-body simulations by \citet{Mayer01} showed that in
general disk-like dwarf irregular galaxies undergo a great
morphological transformation due to close encounters with the more 
massive Milky Way. While dwarfs lose their mass through tidal stripping,
their angular momentum gets gradually removed over a period of 6-10 Gyrs.
This transformation is gradual, and on short timescales 
($\sim2$ Gyrs) a significant angular momentum ($>$80\%) is still preserved. 
Focusing especially on the SMC case, a similar conclusion was 
reached by \citet{Kunkel00} who found that it is extremely hard for 
the SMC to lose its pre-encounter angular momentum during an 
encounter with the LMC.

\subsection{Comparison with HI}

\begin{figure*}
\caption{\label{f:ra_vel1} {\bf [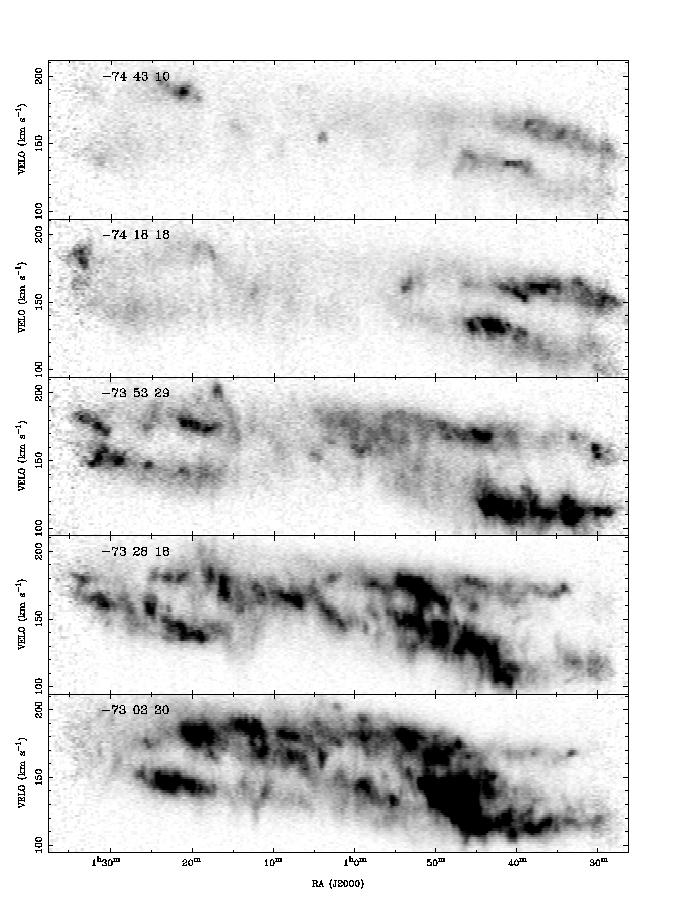]} Ten 
RA-velocity slices through the combined ATCA and
Parkes HI data cube of the SMC, starting with Dec $-74^{\circ} 42' 12''$
and ending with Dec $-72^{\circ} 17' 42''$ (J2000). The grey-scale range 
is: 0--45, 0--45, 0--55, 0--75, 0--75, 0--75, 0--75, 0--55, 0--45, 
0--45 K respectively, with a linear transfer
function.}
\end{figure*}

\setcounter{figure}{6}
\begin{figure*}
\caption{\label{f:ra_vel2} {\bf [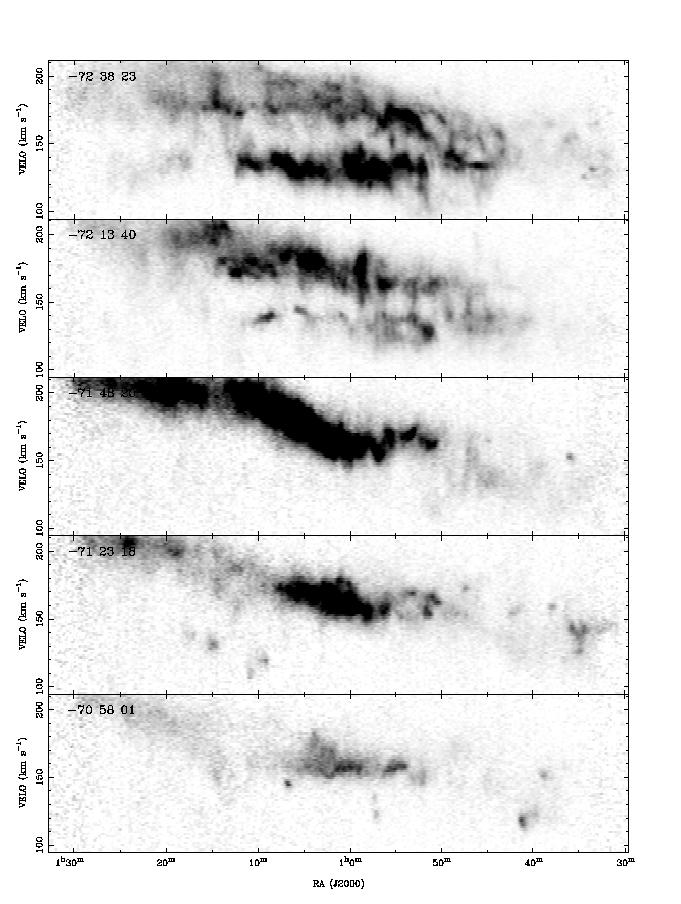]} {\it continued from Fig. 6}}
\end{figure*}

In Section~\ref{s:profile-analysis} we showed that the HI distribution 
contains a large velocity gradient and some signatures for 
the existence of ordered, systematic motions.
This agrees with theoretical simulations which show that the 
current central gaseous body of the SMC  
should still posses angular momentum left over from before the last 
two close encounters with the LMC. 
This angular momentum should be less significant in the case of older 
stellar populations, as indeed has been found from observations 
(see Section~\ref{s:other-tracers} for a summary).
We now compare RA-velocity slices through the HI data cube, shown in
Fig.~\ref{f:ra_vel1}, with the theoretical predictions, shown in Fig. 17 of
\citet{Yoshizawa03} and Fig. 10 of \citet{Gardiner96}.
The area enclosed in our observations contains only the central region of
the SMC ($4.5^{\circ}\times4.5^{\circ}$).

Images in Fig.~\ref{f:ra_vel1} show coherent large-scale features
with a wealth of small-scale structure. 
Many of the features have already been interpreted as expanding shells of gas 
(Staveley-Smith et al. 1997; Stanimirovi\'{c} et al. 1999).
At a low Dec of $\sim-74$\degree, the two most dominant features are 
supergiant shells 37A (RA 01$^{\rm h}$ 25$^{\rm m}$) and 
494A (00$^{\rm h}$ 35$^{\rm m}$). These shells were investigated in  
Stanimirovi\'{c} et al. (1999). Around Dec $\sim-73$\degree 20$'$, 
Fig.~\ref{f:ra_vel1} shows an HI emission, bar-like, feature 
stretching over the RA range 00$^{\rm h}$ 40$^{\rm m}$ to 
$\sim$00$^{\rm h}$ 55$^{\rm m}$, and having a velocity gradient 
from $\sim$120 to $\sim$170 \kms. 
This feature continues eastward as a complex network of filaments 
and shell-like features, culminating around Dec
$\sim-72$\degree 37$'$ with two long filaments centered at 
heliocentric velocities of 130 and 170
\kms. The coherent appearance of these features traced throughout the data
cube was interpreted as the low-Dec hemisphere of the supergiant shell 304A (see
Stanimirovi\'{c} et al. 1999). At a very high Dec of 
$\sim-71$\degree 55$'$, a high-velocity component is apparent, at about 200
\kms.

We do not find obvious features that could correspond to the predicted
eastern and western tails within the region covered by our observations. 
To search for potential tidal tails running out of the central SMC main body
an investigation of a larger area is essential; 
this will be possible in the near future with new observations by
\citet{Bruns02} and \citet{Muller03}, but is beyond the scope of this paper.
Concerning the position and velocity range of the central
component, our observations are in better agreement with the \citet{Gardiner96}
model predictions: the coherent structure seen from 
RA 00$^{\rm h}$ 30$^{\rm m}$ to 01$^{\rm h}$ 30$^{\rm m}$ corresponds to the area
where the central bar-like feature predicted by the model should be found, 
and its velocity span from 100 to 200 \kms~is very close to 
the predicted velocity range for the bar-like component. 
The \citet{Yoshizawa03} model predictions for the left-over disk component cover
a smaller area and velocity range than what is found in the observations.

\section{Rotation curve and mass modeling of the SMC}
\label{s:rotation-curve}

As discussed in the previous section, the central 
$4^{\circ} \times 4^{\circ}$ of the SMC, covered by the 
HI observations, most likely corresponds to the left-over part of the original
SMC's gaseous rotationally supported disk. This disk should contain 
significant signatures of angular momentum left from the pre-encounter disk. 
Motivated by this, we proceed further to derive the rotation curve
from the observed HI velocity field, and the mass
model, assuming that the entire velocity gradient is due to rotation.

\subsection{Tilted ring analysis}

\begin{figure*}
\epsscale{1.} 
\plotone{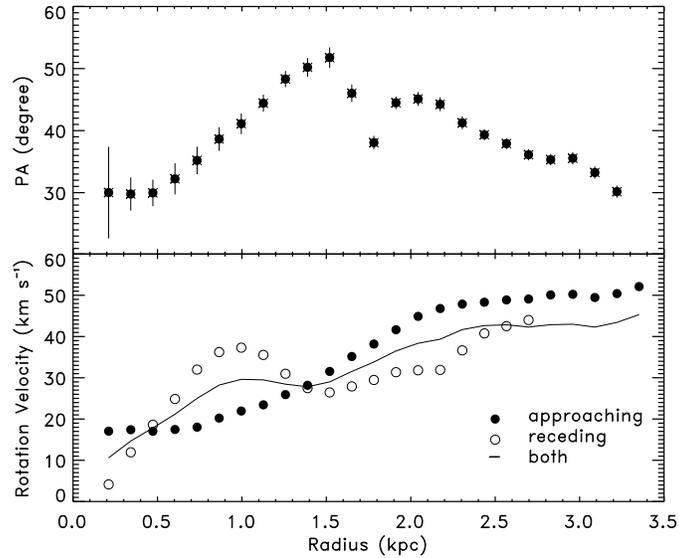}
\caption{\label{f:pa-vrot}The
least-squares solution for the position angle (PA) of the kinematic line of
nodes (top panel) and the rotational velocity ($V_{\rm rot}$) (bottom
panel) as a function of radius. The mean position angle is 40\degree. The
rotation curves for the approaching and receding sides of the velocity
field are shown with filled and open circles, respectively. The global
rotation curve is shown with the solid line.}
\end{figure*}

The velocity field derived in Section~\ref{s:profile-analysis}, and shown in
Fig.~\ref{f:gal_velfield}, was used to define the major and minor 
kinematic axes, and the apparent kinematic
center at $\sim$ RA $01^{\rm h} 05^{\rm m}$, Dec $-72^{\circ} 25'$
(J2000), with a systemic velocity of $\sim$ 20 \kms~in the galactocentric
reference frame. The position angle (PA) of the major kinematic axis is around
50\degree. The tilted ring algorithm {\sc rocur} in the {\sc aips} package 
was then used to derive the HI rotation curve at different 
radii from the center.  
A fairly standard procedure \citep{Meurer98,Bureau02} was applied 
to estimate a set of kinematic parameters that represents well the 
observed velocity field at all radii. It was found that solutions 
for the inclination vary 
greatly especially for the outer radii. In the end, the 
value $i=40$\degree$\pm$20\degree~was adopted. The newly determined 
least-squares solution for the systemic velocity was found to be 24 \kms.
Finally, the solutions for the rotational velocity and position 
angle were determined and are shown in Fig.~\ref{f:pa-vrot}. 
The position angle varies systematically from 30\degree~to 50\degree~(as
shown in the top panel of Fig.~\ref{f:pa-vrot}).  
Using the mean PA, the global and the separate rotation curves for 
the receding and approaching sides of the velocity field were derived, 
as shown in Fig.~\ref{f:pa-vrot} (bottom panel).

The global rotation curve shows a slow rise up to $R\sim3$ kpc, where it
reaches the maximum velocity of $\sim$40 \kms.
The receding and approaching curves are significantly different, which is
not surprising since the velocity field is quite asymmetric
(Fig.~\ref{f:gal_velfield}). The receding curve is less smooth and could have
a local maximum around 1 kpc from the center of rotation.
The maximum rotation velocity found in 
previous studies
\citep{Hindman67,LoiseauBajaja81} was $\sim$ 36 \kms.

The line-of-sight HI velocity dispersion distribution shown in 
Fig.~\ref{f:dispersion-mean} pointed to a high velocity dispersion across
most of the SMC. 
The mean value, $\sigma_{\rm HI}=(22\pm2)$ \kms, is significantly higher than 
what is found for spiral galaxies, or even other dwarf galaxies 
\citep{Meurer98,Cote00}. It is also large compared to the rotational velocity
(Fig.~\ref{f:pa-vrot}). 
If interpreted as being all due to random  motions, rather than bulk
motions along the line-of-sight, such high values for
$\sigma_{\rm HI}$ suggest that  the turbulence in the ISM of the SMC has an
important influence on the system dynamics.  

To account for this dynamical support we estimated the 
asymmetric drift correction. 
The usual prescription for determining this correction \citep{Meurer96}
was followed:
we derived the azimuthally averaged HI surface brightness and velocity
dispersion profiles (for deprojected
circular annuli with a mean position angle of 40\degree~and an inclination of
40\degree), and assumed a constant vertical scale height.
In general, the derived asymmetric drift correction is quite significant and range 
from 0, for the inner radii, to $\sim 40$ \kms~for the outer radii. 
The observed rotation curve corrected for the pressure
support is presented in Fig.~\ref{f:rotmod-final} (crosses). The difference is
significant for $R>0.5$ kpc with the corrected curve being higher
by $\sim$ 10 \kms~than the observed one.

\subsection{The mass model}

\begin{figure*}
\epsscale{1.}
\plotone{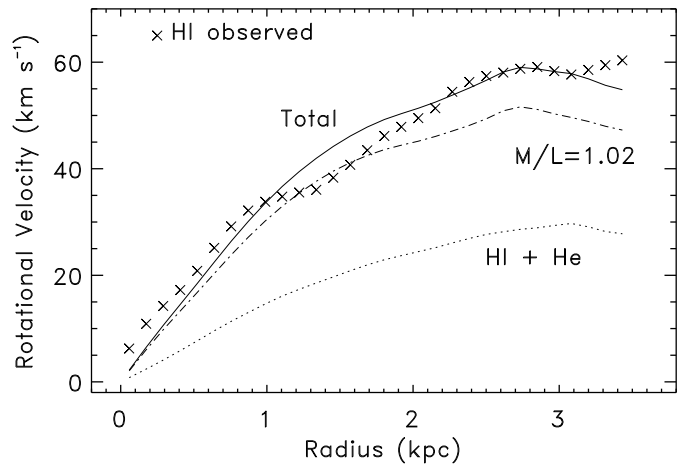}
\caption{\label{f:rotmod-final} The observed rotation curve, corrected for the
pressure support (crosses), compared with the total predicted rotation
curve (solid line), composed of: (a) the gaseous, HI$+$He, component
(dotted line); (b) the stellar component (dash-dotted line) with an estimated
mass-to-light ratio $M_{\ast}/L_{\rm V}=1.02$.}
\end{figure*}

A two-component mass model was fitted to the corrected rotation curve.

(a) To derive the deprojected rotation curve due to the potential resulting
from the neutral gas alone, the radial HI surface density profile
was used in the {\sc gipsy}'s {\sc rotmod} task \citep{gipsy}.  
An exponential density law was assumed for the vertical disk distribution, 
with a scale height of 1 kpc. 
An indication of the large scale height was given by the typical size 
of large HI shells in the SMC, as well as from previous estimates 
based on the velocity dispersion and an average surface density of matter
for the case of an isothermal disk.
The resultant rotation curve ($V_{\rm g}$) is shown in
Fig.~\ref{f:rotmod-final} with dotted line.

(b) To derive the deprojected rotation curve due to the optical surface
density distribution, the V-band image (shown in Fig.~\ref{f:v-HI}) 
was used. This image was first smoothed to
20 arcmin resolution and scaled to match the total luminosity of the SMC in
the V-band of $3.1\times10^{8}$ L$_{\odot}$ \citep{RC3}. 
The rotation curve arising from the stellar potential alone ($V_{\ast}$)
was estimated from the V-band stellar density profile assuming, 
in first approximation, that $M_{\ast}/L_{\rm V}=1$ (see the curve shown as a 
dash-dotted line in Fig.~\ref{f:rotmod-final}).  
The best fit of the
total predicted rotational velocity, $\sqrt{V_{\rm g}^{2}+V_{\ast}^{2}}$, to
the observed rotational velocity corrected for the pressure support was
obtained for $M_{\ast}/L_{\rm V}=1.02$ (solid line in Fig.~\ref{f:rotmod-final}).
The two component mass model appears to fit the observed rotational
velocities quite well, suggesting that no additional component, such as a
dark halo, is needed to explain the rotation of the SMC.

The rotation velocity of only the stellar component implies a total
stellar mass of the SMC of $1.8\times10^{9}$ M$_{\odot}$, within a
radius of 3 kpc. From Section~\ref{s:hi-data} and after correction for
neutral He, the mass of HI$+$He is $5.6\times10^{8}$ M$_{\odot}$, being
almost a third of the stellar mass within the same radius. The total mass
of the SMC, implied from the rotation curve, is thus $2.4\times10^{9}$
M$_{\odot}$. This is almost twice that of $1.5\times10^{9}$ M$_{\odot}$,
derived by \citet{Hindman67} within the slightly smaller radius of 2.6
kpc, and is similar to the inital mass of the SMC assumed by
\citet{Gardiner94} (about $2\times10^{9}$ M$_{\odot}$) and by
\citet{Yoshizawa03} ($3\times10^{9}$ M$_{\odot}$).

\subsubsection{The stellar $M/L$ from population synthesis models}

Stellar population synthesis (SPS) models provide an independent way to 
estimate the stellar mass-to-light ratio. \citet{Bell01} used simplified 
spiral galaxy evolution models, based on several different SPS models, 
to investigate trends of the stellar $M/L$ with galaxy
properties, such as colors and gas fraction. 
They found good agreement with observational data and showed that
the trend of $M/L$ with color, 
$\log_{10} (M/L)=a_{\lambda}+b_{\lambda}$Color, is robust with respect 
to the SPS models but is dependent on the type of the initial mass 
function (IMF) used. They tabulate coefficients $a_{\lambda}$ and 
$a_{\lambda}$ for different colors, stellar metallicity, SPS models, and IMFs.

We have used coefficients for the case of a stellar metallicity appropriate
for the SMC, which is about 25\% of the solar
abundance,  based on the SPS model by \cite{Bruzual03}, for 
a Salpeter IMF scaled by a factor of 0.7 \citep{Bell01}, and
the \cite{Schmidt59} star formation law. The stellar $M/L_{\rm V}$ was then
estimated using the broad band colors given in Table 1, $B-V=0.41$ and $B-R=0.7$.
We get $M/L_{\rm V}=0.8\pm0.2$ which agrees with the value obtained 
from the HI rotation curve.

\subsection{Critical density for star formation}

\begin{figure*}
\epsscale{1.}
\plotone{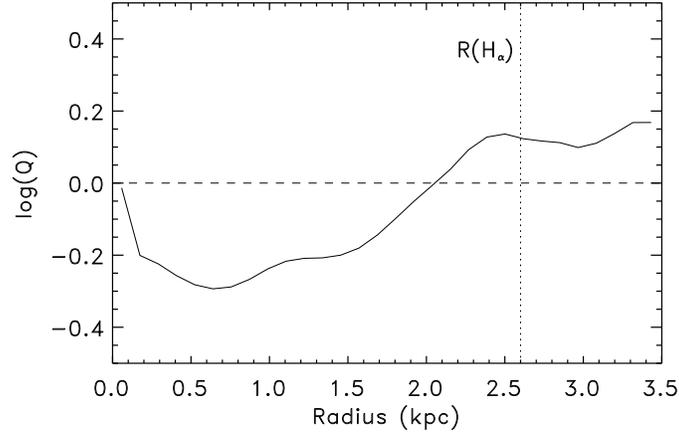}
\caption[Radial variation of Toomre's disk stability 
parameter $Q$.]{\label{f:Q} Radial variation of Toomre's disk stability
parameter $Q$. The dashed line shows $Q=1$ which separates disk-stable
($Q>1$) from disk-unstable phase ($Q<1$). 
The radius where the H$\alpha$ profile departs from an exponential
distribution, at approximately $R({\rm H}\alpha)=2.6$ kpc, 
is shown with a dotted line.}
\end{figure*}

\begin{figure*}
\epsscale{1.}
\plotone{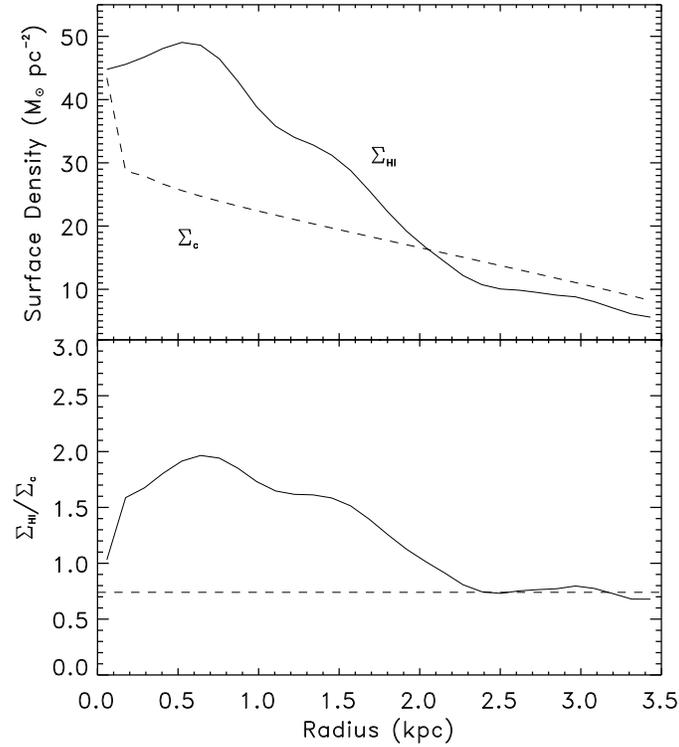}
\caption[The 
observed and critical surface density.]{\label{f:stability} Top panel: the
radial distribution of the observed ($\Sigma_{\rm HI}$) and critical
($\Sigma_{\rm c}$) surface density of the gas in the SMC. Bottom panel: the
radial distribution of the ratio of the observed to the critical surface
density ($\alpha=
\Sigma_{\rm HI}/\Sigma_{\rm c}$). The dashed line shows $\alpha=0.74$.}
\end{figure*}

We now go a step further and address the gravitational stability of the
current gaseous disk of the SMC and its possible consequences for star
formation. We estimated the disk stability parameter, $Q$, \citep{Toomre64}
using the mean velocity dispersion, azimuthally averaged HI surface
brightness, and the angular velocity derived using the rotation velocity.
According to the Toomre's stability criterion, the disk is stable 
for $Q>1$, while regions with $Q \la 1$ are unstable and under 
possible fragmentation which can lead to new star formation 
\citep{BinneyTremaine}. The estimated values of $Q$, as a function of
radius, are shown in Fig.~\ref{f:Q}. 
The plot shows that most of the gaseous disk, for
$R<2.1$ kpc, is unstable and, most likely, under current star
formation. The unstable region appears to be quite extended, reaching
almost the cut-off H$\alpha$ radius, $R({\rm H}_{\alpha})\approx2.6$
kpc (based on H$\alpha$ observations by \cite{Kennicuttetal95}), 
which delineates the parts of 
the gaseous disk under the most recent star formation.

We also determined the so called critical surface
density, defined as the surface density needed to stabilize the
disk (or for $Q=1$). Both the observed
($\Sigma_{\rm HI}$) and critical ($\Sigma_{\rm c}$) surface densities are
shown in Fig.~\ref{f:stability} (top panel). 
The ratio of the observed to critical surface density 
($\alpha=\Sigma_{\rm HI}/\Sigma_{\rm c}=1/Q$) is shown in the
same figure (bottom panel). For the unstable part of the disk, $\alpha$
varies between 1.0 and 2.0, while it has a constant value of
$\alpha=0.74\pm0.04$ for the stable part of the disk. 
This agrees with the empirical relationship between the star formation 
threshold and the disk stability found by \citet{Kennicutt89} for a sample of
normal galaxies, whereby most of the star formation is taking
place for $\alpha>0.67$ (and hence for $Q<1.5$).

\section{Summary and concluding remarks}
\label{s:summary}

We have used the latest HI observations, obtained with the ATCA and the
Parkes telescope, to re-examine the kinematics of the SMC.
The HI velocity field, derived in Section~\ref{s:profile-analysis},
shows a large velocity gradient from the south-west to the north-east. The
iso-velocity contours of this velocity field show some symmetry,
suggestive of a differential rotation. 
Some large-scale distortions in this velocity field are easily visible, but
could be related to positions of several supergiant shells.
In contrast to the HI distribution, the old stellar
populations appear to have a spheroidal spatial distribution and 
a total absence of rotation.
In Section~\ref{s:3D-structure}, we summarized previous observational models
of the SMC and the greatly controversial issue of its possibly 
large depth along the line-of-sight. 
We cautioned that the differential extinction across the SMC,
revealed recently by \citet{Zaritsky02}, could be a significant source, 
unappreciated previously, for an overestimate of stellar distances, and
therefore the SMC's depth. 

In Section~\ref{s:theoretical-models}, we summarized several tidal models
that are concerned with the 3-D structure and kinematics of the SMC.
In order to reproduce observational characteristics of the Magellanic
System these models needed to assume the existence of angular momentum in the
pre-encounter SMC (about 1.5--2 Gyr ago). Models predict that the current,
left-over material from the pre-encounter SMC disk is a disk-like or a 
bar-like feature,
about 4 kpc in extent, with a velocity gradient of about 100
\kms. This velocity gradient is at least partially
from the original angular momentum, as it is very difficult to lose the 
original spin during a galaxy encounter \citep{Mayer01,Kunkel00}.
The elongation along the line-of-sight of
the pre-encounter SMC disk may also be partially responsible.
It is, however, not clear from the models what are the relative contributions of 
these two effects to the predicted velocity gradient.
In addition, theoretical models predict higher velocity dispersion for the
post-encounter stellar disk without a significant velocity gradient; this
agrees with observations of older stellar populations summarized in 
Section~\ref{s:other-tracers-kinematics}.

While we do not find evidence for the existence of tidal tails
predicted by the models, the central region of the SMC covered in 
these observations most likely corresponds to the central 
component left-over from the pre-encounter SMC gaseous 
(rotationally supported) disk. 
The observed HI velocity gradient agrees well with the model predictions. 
We then proceeded further to derive its rotation curve and mass
model in Section~\ref{s:rotation-curve}.
A set of kinematic parameters derived from the tilted ring analyses agrees
extremely well with the assumptions used in theoretical models that led to a good
reproduction of observational properties of the Magellanic System. 
The derived HI rotation curve in Section~\ref{s:rotation-curve} 
rapidly rises to about 60 \kms~up to the
turnover radius of $\sim3$ kpc. A stellar mass-to-light ratio of about
unity was required to scale the stellar component of the SMC's rotation 
curve to match the observed rotation curve. This suggests that a
dark matter halo is not needed to explain the dynamics of the SMC. 
This is a surprising result, as dwarf irregular galaxies are
often found to be dark matter dominated, and could be related to the
mechanism of tidal stripping, as indicated by \cite{Mayer01}. The total
dynamical mass of the SMC derived from the rotation curve is $2.4\times10^{9}$
M$_{\odot}$, three quarters of which are due to the stellar mass only.
We also derived Toomre's disk stability parameter $Q$, which shows 
that almost all of the SMC disk is in the unstable regime. The
gravitationally stable part, on the other hand, has a ratio of the
observed to critical surface density $\alpha=0.74$ which is in excellent
agreement with the  empirical stability threshold found by
\citet{Kennicutt89}.

All of the above suggest that the HI distribution is capable of providing
valuable information about the SMC dynamics. Although very disturbed, it
still contains imprints of the original system prior to 
the latest encounters with the LMC and the Galaxy. 
The HI data discussed here comprise only the central
$4.5^{\circ}\times4.5^{\circ}$ of the SMC. 
It is very important to compare tidal theoretical models
with data comprising a larger area around the SMC to search for
signatures of the proposed tidal tails. This will be
possible in the near future using the Parkes Multibeam HI
observations of the whole Magellanic System \citep{Bruns00}, as well as the
ATCA and Parkes observations of the Magellanic Bridge \citep{Muller03}.
 
The HI SMC data set is available from the ATNF SMC web page \\
(http://www.atnf.csiro.au/research/smc\_h1/).

\section*{Acknowledgments}

We are greatly thankful to Mike Bessell for providing us with the V-band optical
image of the SMC prior to publication. 
We thank Greg Bothun for sending us the H$\alpha$ image.
We thank John Dickey, Jacco van Loon, and Mary Putman for 
stimulating and fruitful discussions, and an anonymous referee for
insightful comments.  
This work was supported in part by NSF grants  AST-0097417 and AST-9981308.

\label{lastpage}
\end{document}